\documentclass{article}
\usepackage{spconf,amsmath,graphicx}
\usepackage{algpseudocode}
\usepackage{algorithm}
\usepackage{amsfonts}
\usepackage{booktabs}
\usepackage{color}
\usepackage[colorlinks=true]{hyperref}
\usepackage{movie15}
\usepackage{multirow}
\usepackage{subcaption}
\usepackage{tikz}

\newcommand{\Circle}[1]{\tikz[baseline]{\small \node[anchor=base, draw, circle, inner sep=0, minimum width=0.95em]{#1};}}
\newcommand{\FigDog}[2]{
  \begin{subfigure}{0.158\textwidth}
  \includemovie[poster=images/#1.png,mouse,repeat]{1\linewidth}{.3\linewidth}{audios/#1.mp3}
  \caption{#2}
  \label{fig:#1_sample}
\end{subfigure}}
\newcommand{\FigDigit}[1]{
  \begin{subfigure}{0.044\textwidth}
  \includemovie[poster=images/#1.png,mouse,repeat]{1\linewidth}{.6\linewidth}{audios/#1.mp3}
  \caption{\texttt{#1}}
  \label{fig:#1}
\end{subfigure}}

\title{Feature-Rich Audio Model Inversion for Data-Free Knowledge Distillation Towards General Sound Classification}

\name{Zuheng Kang$^*$, Yayun He$^*$\thanks{$^*$Equal contributions}, Jianzong Wang, Junqing Peng, Xiaoyang Qu, Jing Xiao$^\dagger$\thanks{ $^\dagger$Corresponding author: Jing Xiao, xiaojing661@pingan.com.cn}}
\address{Ping An Technology (Shenzhen) Co., Ltd.}

\begin{document}

\maketitle

\begin{abstract}
  Data-Free Knowledge Distillation (DFKD) has recently attracted growing attention in the academic community, especially with major breakthroughs in computer vision.
  Despite promising results, the technique has not been well applied to audio and signal processing.
  Due to the variable duration of audio signals, it has its own unique way of modeling.
  In this work, we propose feature-rich audio model inversion (FRAMI), a data-free knowledge distillation framework for general sound classification tasks.
  It first generates high-quality and feature-rich Mel-spectrograms through a feature-invariant contrastive loss.
  Then, the hidden states before and after the statistics pooling layer are reused when knowledge distillation is performed on these feature-rich samples.
  Experimental results on the Urbansound8k, ESC-50, and audioMNIST datasets demonstrate that FRAMI can generate feature-rich samples.
  Meanwhile, the accuracy of the student model is further improved by reusing the hidden state and significantly outperforms the baseline method.
\end{abstract}


\section{Introduction}
\label{sec:intro}

The goal of Knowledge Distillation (KD) is to assist a less parameterized student model to achieve similar generalization capabilities by training data to mimic the behavior of a powerful teacher model with a large number of parameters \cite{hinton2015distilling}.
Due to its simple and efficient design, this has led to the great success of KD technology in various audio applications, such as audio classification \cite{choi2022temporal}, automatic speech recognition \cite{kurata2020knowledge}.
In addition to aligning category predictions under the same inputs, some researchers have constrained additional hidden state information from intermediate teacher-student layer pairs to better train student models \cite{choi2022temporal,jiao2019tinybert,chen2022knowledge}.
However, it is highly dependent on the particular design of the network structure with similar knowledge representation.

Due to confidentiality or privacy issues, in many cases, the raw data is not available and only pre-trained models can be used.
Data-Free Knowledge Distillation (DFKD) enables KD of student models on data generated through model inversion \cite{yin2020dreaming,luo2020large}.
It has been successfully applied in the fields of natural language processing \cite{ma2020adversarial}, federated learning \cite{zhu2021data}, computer vision (CV) \cite{chen2019data,zhang2021data,chawla2021data,hu2022data}, etc.
The KD accuracy of the student model can also be improved by increasing the diversity of the generated samples with adversarial learning \cite{fang2019data,qu2021enhancing}, contrastive learning \cite{fang2021contrastive} or other methods \cite{fang2022up,Li2020LearningIS}.
Nevertheless, none of these methods are designed for audio.

KD in general sound classification is also affected by sensitive data or copyright issues.
In scenarios such as smart homes, environmental monitoring, and mechanical fault detection, models usually need to be deployed on the edge.
But these training data are difficult to obtain, making it impossible to perform KD.
Hence, DFKD is quite necessary.
In the past decades, general sound classification has often been based on hand-crafted features, such as Mel-spectrograms, and using statistics pooling (SP) with classifiers as back-end models.
However, since audio features are usually long and temporally rich, it is difficult to produce high-quality spectrograms using traditional model inversion methods.
Meanwhile, the SP drops the temporal information, resulting in the underutilization of the temporally hidden state information in the KD.

This work attempts to address the above issues and focuses on investigating a model inversion approach to generate high-quality and feature-rich samples.
Since the samples generated by the traditional method have very sparse effective features in the time dimension, this results in the student model learning only a very limited amount of knowledge from the teacher model during the KD process.
We exploit the feature invariance of the samples in the time dimension to generate stable and feature-rich samples.
These samples are then used to simultaneously learn temporal feature-rich hidden state information during the KD stage to achieve higher accuracy of the student model.

The main contribution is as follows:
\Circle{1}
This paper proposed the feature-rich audio model inversion (FRAMI), which first uses feature-invariant contrastive loss to ensure the generation of feature-rich samples.
\Circle{2}
The hidden state before and after the statistics pooling layer is then reused to further improve student performance in KD.
\Circle{3}
Extensive experiments verify the effectiveness of the above two methods for Urbansound8k, ESC-50, and audioMNIST and provide reasonable explanations.
Also, the generated samples are clearly audible.

\section{Background Knowledge}
\label{sec:motivation}

\subsection{Data-free Knowledge Distillation}
\label{subsec:dfkd}

Excellent data-free knowledge distillation (DFKD) usually results from the generation of high-quality and feature-rich data.
Model inversion, as an important step in DFKD, aims to recover the possible training data $ \mathcal{X}' $ as realistically as possible from the pre-trained teacher model $ f_t\left( x;\theta _t \right) $ to replace the inaccessible original data $ \mathcal{X} $.
The student model $ f_s\left(x;\theta_s\right) $ can then be trained using these datasets $ \mathcal{X}' $ to mimic the teacher's behavior without acquiring the original data $ \mathcal{X} $.
Although model inversion can produce near-realistic data, a lack of data diversity may prevent student models from learning as much as possible from teacher models through KD.
Methods such as adversarial learning \cite{fang2019data,qu2021enhancing} and contrastive learning \cite{fang2021contrastive} can produce the richest possible samples and cover as much of the input space as possible.

\vspace{-0.4em}
\subsection{Motivation}
\label{subsec:motivation}

Although DFKD has achieved remarkable results in the field of computer vision (CV), it has not made many breakthroughs in the field of audio and signal processing.
The dimensionality of images is fixed.
In contrast, in audio modeling, the temporal dimension is not fixed and may be long, and researchers often use methods such as statistics pooling (SP) to eliminate variable temporal dimensions.
In a traditional model inversion, when large spectra are generated, the effective feature distribution in the time dimension is usually very sparse.
That is, the presence of only a few features in a short period of time is sufficient to determine its category.
Therefore, we need to find a model inversion method to guarantee the feature richness of the generated audio samples in the time dimension.
Meanwhile, it is also necessary to design a mechanism that can effectively utilize these feature-rich samples in KD to improve the recognition accuracy.

\section{Methodology}
\label{sec:method}

\subsection{Feature Invariance Contrastive Inversion}
\label{subsec:fic}

\begin{figure}[t]
  \includegraphics[width=0.5\textwidth]{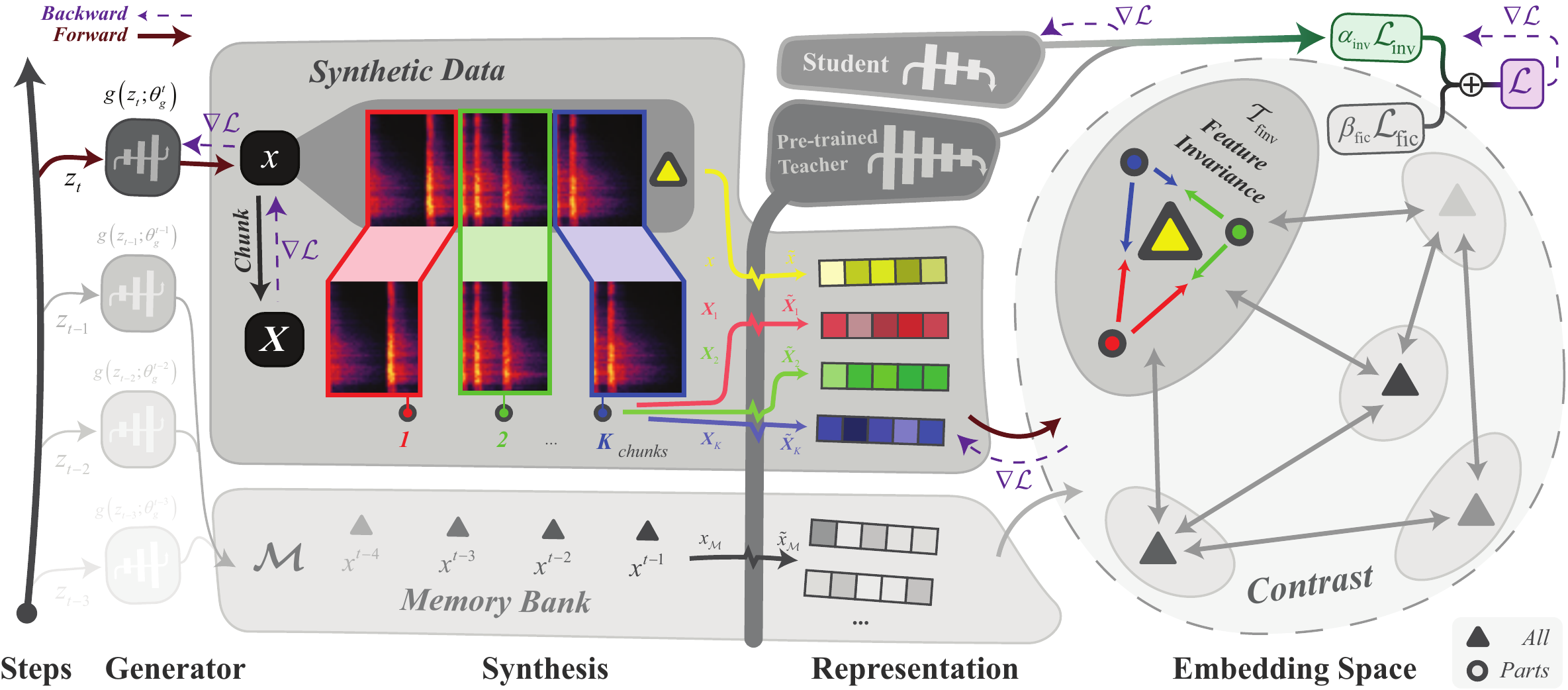}
  \centering
  \caption{Feature invariance contrastive inversion overview.}
  \label{fig:finv}
  \vspace{-1em}
\end{figure}

\textbf{Feature Invariance:}
Audio signals are either time-dependent (TD) or time-independent (TID).
For example, voice command is TD because different information exists at different times.
Environmental sound is TID because it is usually stable in the time dimension.
Namely, for each time period, it needs to be considered as belonging to the same category.
To ensure that the features of each generated part should be consistent, we propose a technique called feature invariance to handle the TID case, shown in Fig. \ref{fig:finv}.
That is, when performing model inversion, it is assumed that the generated samples are $ x=\left\{ x_1,x_2,...,x_T \right\} $, we can randomly divide it into several chunks along time dimension, denoted as the set $ \boldsymbol{X} $ in Eq. \ref{eq:fi_chunk}, where $ K $ is a randomly selected from the range under $ K\in \left[ K_{\min},K_{\max} \right] $.

\noindent
\begin{equation}
  \small
  \label{eq:fi_chunk}
  \boldsymbol{X} =\underset{K\,\,\mathrm{chunks}}{\underbrace{\left[ \left[ x_1,...,x_k \right] ,\left[ x_{k+1},...,x_{2k} \right] ,...,\left[ ...,x_T \right] \right] }}
\end{equation}
\noindent

To show that the features are consistent across different chunks, we first transform each sample $ x $ into a latent feature embedding space $ \tilde{x} $ with the projection function $ h $, which is an instance discriminator for sample $ x $ extracted from the teacher network $ f_t $ \cite{fang2021contrastive}.
Due to the feature invariance of TID audio, the distance between these feature embeddings should be close.
Regarding the measure of distance, we can use a simple cosine distance to represent the distance between two features $ \tilde{x}_i $ and $ \tilde{x}_j $, denoted as, $ \mathrm{sim}\left( \tilde{x}_i, \tilde{x}_j \right) $.
Then, the feature invariant term $ \mathcal{T} _{\mathrm{finv}} $ is defined in Eq. \ref{eq:finv_term} to all possible $ \left( \tilde{x}_i,\tilde{x}_j \right) $ pairs in $ \tilde{\boldsymbol{X}} $.

\noindent
\begin{equation}
  \small
  \label{eq:finv_term}
  \mathcal{T} _{\mathrm{finv}}\left( \tilde{\boldsymbol{X}} \right) =\mathbb{E} _{\left( \tilde{x}_i,\tilde{x}_j\in \tilde{\boldsymbol{X}} \right) \land \left( i < j \right)}\left[ \mathrm{sim}\left( \tilde{x}_i,\tilde{x}_j \right) \right]
\end{equation}
\noindent

\noindent
\textbf{Contrastive Learning:}
It is also necessary to guarantee the diversity of the generated features.
For each newly generated sample $ x \in \mathcal{X} ' $, we build a positive sample $ x^+ $ by randomly adding audio data augmentation, and consider other samples $ x^- $ as negative samples.
Since the split audio is also intercepted from the original audio, it can also be considered as a positive sample.
The feature invariance contrastive loss $ \mathcal{L} _{\mathrm{fic}} $ can be defined in Eq. \ref{eq:contrastive}, where $ \tau $ is the temperature.

\noindent
\begin{equation}
  \footnotesize
  \label{eq:contrastive}
  \hspace{-1mm}
  \mathcal{L} _{\mathrm{fic}}\left( \mathcal{X}' \right) =-\mathbb{E} _{x_i\in \mathcal{X}'}\left[ \log \frac{\exp \left( \left( \mathrm{sim}\left( \tilde{x}_i,\tilde{x}_{i}^{+} \right) +\mathcal{T} _{\mathrm{finv}}\left( \tilde{\boldsymbol{X}}_i \right) \right) /\tau \right)}{\sum_j{\exp \left( \mathrm{sim}\left( \tilde{x}_i,\tilde{x}_{j}^{-} \right) /\tau \right)}} \right]
\end{equation}
\noindent

\noindent
\textbf{Inversion Loss:}
In addition to the above loss, we also need to add deep inversion loss $ \mathcal{L}_{\mathrm{inv}} $ in Eq. \ref{eq:inv_loss}, \cite{yin2020dreaming}: a class confidence loss $ \mathcal{L}_{\mathrm{cls}} $, an adversarial loss $ \mathcal{L}_{\mathrm{adv}} $ and a feature regularization loss $ \mathcal{L}_{\mathrm{bn}} $ in Eq. \ref{eq:other_loss}, where $ \alpha $, $ \beta $ and $ \gamma $ are hyperparameters.

\noindent
\begin{equation}
  \small
  \label{eq:other_loss}
  \begin{cases}
    \mathcal{L} _{\mathrm{cls}}=\mathrm{CrossEntropy}\left( f_t\left( x \right) ,c \right)                                                                                                          \\
    \mathcal{L} _{\mathrm{adv}}=-\mathrm{KLD}\left( f_t\left( x \right) /\tau \parallel f_s\left( x \right) /\tau \right)                                                                           \\
    \mathcal{L} _{\mathrm{bn}}=\sum_l{\left( \left\| \mu _{\mathrm{feat}}^{l}-\mu _{\mathrm{bn}}^{l} \right\| _2+\left\| \sigma _{\mathrm{feat}}^{l}-\sigma _{\mathrm{bn}}^{l} \right\| _2 \right)} \\
  \end{cases}
\end{equation}
\noindent

\noindent
\begin{equation}
  \small
  \label{eq:inv_loss}
  \mathcal{L} _{\mathrm{inv}}=\alpha \cdot \mathcal{L} _{\mathrm{bn}}+\beta \cdot \mathcal{L} _{\mathrm{cls}}+\gamma \cdot \mathcal{L} _{\mathrm{adv}}
\end{equation}
\noindent

The detailed procedure of feature invariant contrastive inversion is shown in the pseudo-code of the Algorithm \ref{pseudo:fic}.

\begin{algorithm}[t]
  \floatname{algorithm}{Algorithm}
  \renewcommand{\algorithmicrequire}{\textbf{Input:}}
  \renewcommand{\algorithmicensure}{\textbf{Output:}}
  \caption{FRAMI Training Policy}
  \label{pseudo:fic}
  \footnotesize
  \begin{algorithmic}[1]
    \Require A pre-trained teacher $ f_t\left( \cdot;\theta _t \right) $
    \Ensure A spectral memory bank $ \mathcal{M} $
    \State $ \mathcal{M} \gets \phi $
    \State initialize student $ f_s\left( \cdot;\theta _s \right) $
    \State initialize discriminator $ h\left( \cdot;\theta _h \right) $
    \For{$ epoch\in \left[ 1,end \right] $}
    \State \textcolor[rgb]{0.6,0.6,0.6}{/* model inversion phase */}
    \State initialize generator $ g\left( \cdot;\theta _g \right) $
    \State $ z\gets \mathcal{N} \left( 0,1 \right) $
    \For{$ i^{th} $ update steps }
    \State $ x_i\gets g\left( z;\theta _g \right) $
    \State $ \boldsymbol{X}_i\gets \mathrm{chunk}\left( x_i,K \right) $
    \State $ x_{\mathrm{M}}\gets \mathrm{sample}\left( \mathcal{M} \right) $
    \State get feature invariance contrastive loss $ \mathcal{L} _{\mathrm{fic}} $ with Eq. \ref{eq:contrastive}
    \State get inversion loss $ \mathcal{L} _{\mathrm{inv}} $ with Eq. \ref{eq:inv_loss}
    \State $ \mathcal{L} \gets \alpha _{\mathrm{fic}}\cdot \mathcal{L} _{\mathrm{fic}}\left( x_{\mathrm{M}}\cup x_i; \boldsymbol{X}_i \right) +\beta _{\mathrm{inv}}\cdot \mathcal{L} _{\mathrm{inv}}\left( x_i \right) $
    \State create and apply gradients with optimizer:
    \State \hspace{\algorithmicindent} update model parameters $ \theta_g $, $ \theta_s $ and $ \theta_h $
    \EndFor
    \State $ \mathcal{M} \gets \mathcal{M} \cup \underset{\mathcal{L}}{\mathrm{argmin}}\left( x \right) $
    \State \textcolor[rgb]{0.6,0.6,0.6}{/* knowledge distillation phase */}
    \State train student model $ f_s $ with $ \mathcal{M} $ in $ \S $ \ref{subsec:reusedKD}
    \EndFor
    \State \Return $ \mathcal{M} $
  \end{algorithmic}
\end{algorithm}

\subsection{Reused Teacher Backend Knowledge Distillation}
\label{subsec:reusedKD}

\begin{figure}[t]
  \includegraphics[width=0.50\textwidth]{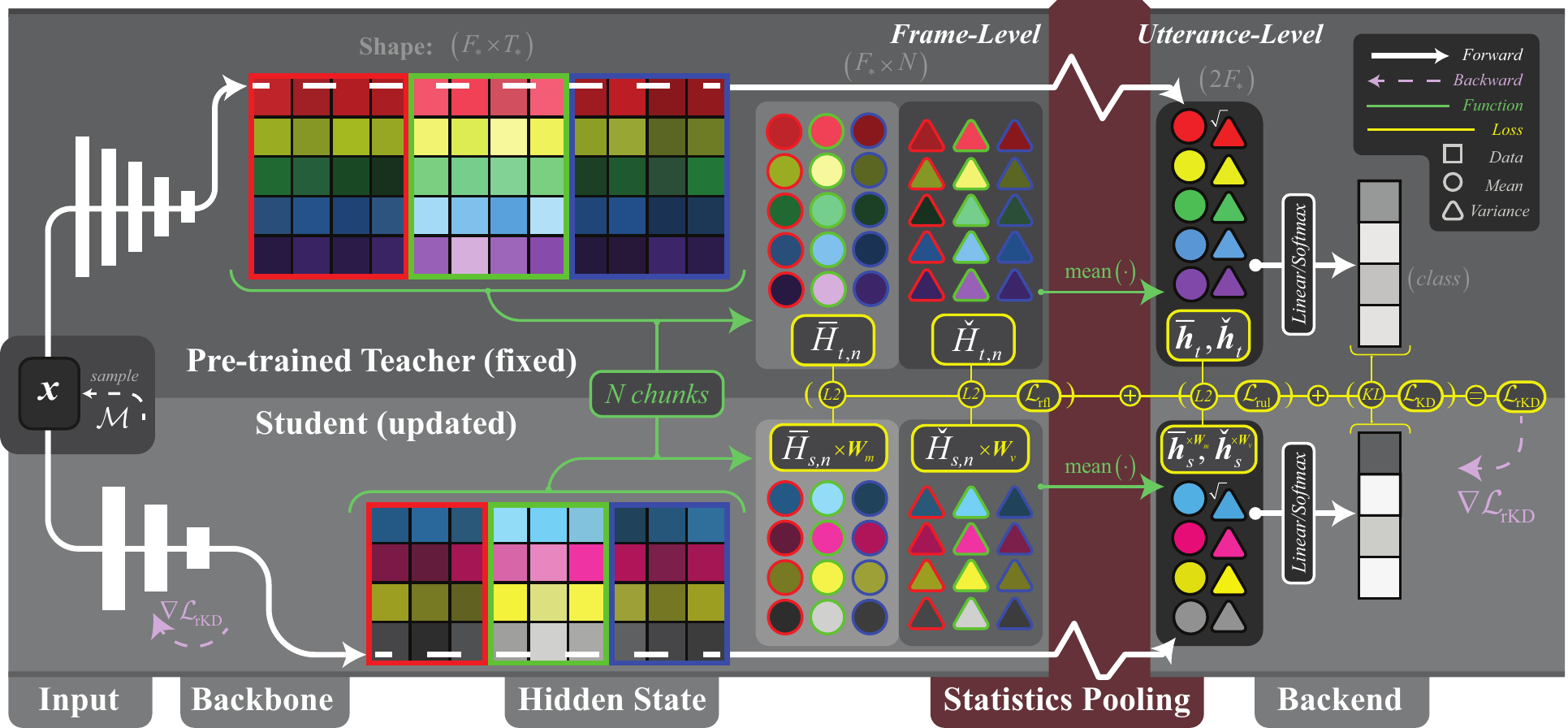}
  \centering
  \caption{Reused teacher backend KD overview.}
  \label{fig:reused}
  \vspace{-1em}
\end{figure}

Recently, various DFKD frameworks have been proposed, but almost all of them use only soft classification logits for KD.
In general sound classification tasks, backend models typically use a statistics pooling (SP) with mean and standard deviation \cite{snyder2018x,qu2020evolutionary}: before SP is the frame-level, and after SP is the utterance-level.
Assuming that the teacher model performs well.
When the student model imitates the teacher model, it will be more robust if the student model can simultaneously learn the additional hidden states before and after SP.
The overall architecture is shown in Fig. \ref{fig:reused}.

\noindent
\textbf{Reused Frame-Level Hidden State:}
Suppose $ h_t\in \mathbb{R} ^{F_t\times T_t} $ and $ h_s\in \mathbb{R} ^{F_s\times T_s} $ are the hidden state of the teacher and student models before SP.
Since they have different dimensions, if there exists a simple mapping that translates the hidden state of the student model into that of the teacher model, then the student model can also learn the hidden information from the teacher model.
To solve the problem of two hidden states having different sizes, there is an \textit{Assumption for Spectral}:
\Circle{1}
No matter how the model scales the hidden states, the order of the information in the time dimension does not change.
\Circle{2}
There may be a simple mapping relationship in the frequency dimension.
Based on this assumption, we can consider the time dimension of both hidden states of $ h_t $ and $ h_s $ as $ N $ chunks in Eq. \ref{eq:hs_chunk}, where $ * $ is either $ t $ or $ s $, $ N $ is a randomly selected from the range under $ N\in \left[ N_{\min},N_{\max} \right] $.

\noindent
\begin{equation}
  \small
  \label{eq:hs_chunk}
  \boldsymbol{H}_* =\underset{N\,\,\mathrm{chunks}}{\underbrace{\left[ \left[ h_{*,1},...,h_{*,n} \right] ,\left[ h_{*,n+1},h_{*,2n} \right] ...,\left[ ...,h_{*,T_*} \right] \right] }}
\end{equation}
\noindent

For the mean, we can average each chunk along time dimension with $ \bar{\boldsymbol{H}}_* \in \mathbb{R} ^{F_*\times N} $.
For the variance, assuming $ \bar{\boldsymbol{h}}_* $ and $ \check{\boldsymbol{h}}_* $ are the overall mean and variance of $ \boldsymbol{H}_* $, we can find a pseudo-variance for chunk $ n $ along time dimension with $ \check{\boldsymbol{H}}_* \in \mathbb{R} ^{F_*\times N} $ in Eq. \ref{eq:pseudo_var}.

\noindent
\begin{equation}
  \small
  \label{eq:pseudo_var}
  \check{\boldsymbol{H}}_{*,n}=\frac{1}{T_*}\sum_{i=n\cdot \left( T_*/N \right) +1}^{\left( n+1 \right) \cdot \left( T_*/N \right)}{\left( h_{*,i}-\bar{\boldsymbol{h}}_* \right) ^2}
\end{equation}
\noindent

Suppose there exist two projections (affine transformations) in the frequency dimension that can transform the mean and variance of the student's hidden state to the teacher's hidden state respectively, with $ \bar{\boldsymbol{H}}_{t,n} = \boldsymbol{W}_m \bar{\boldsymbol{H}}_{s,n} $ and $ \check{\boldsymbol{H}}_{t,n} = \boldsymbol{W}_v \check{\boldsymbol{H}}_{s,n} $, where $ \boldsymbol{W}_m $, $ \boldsymbol{W}_v \in \mathbb{R} ^{F_t\times F_s} $.
Then, the resued frame-level loss $ \mathcal{L}_{\mathrm{rfl}} $ can be defined in Eq. \ref{eq:rfl_loss}, where $ \mathrm{MSE}\left( \cdot \right) $ is the mean square error operator.

\noindent
\begin{equation}
  \small
  \label{eq:rfl_loss}
  \mathcal{L} _{\mathrm{rfl}}=\mathbb{E} _{n\in N}\left[ \begin{array}{l}
      \mathrm{MSE}\left( \boldsymbol{W}_m\bar{\boldsymbol{H}}_{s,n},\bar{\boldsymbol{H}}_{t,n} \right)      \\
      +\mathrm{MSE}\left( \boldsymbol{W}_v\check{\boldsymbol{H}}_{s,n},\check{\boldsymbol{H}}_{t,n} \right) \\
    \end{array} \right]
\end{equation}
\noindent

\noindent
\textbf{Reused Utterance-Level Hidden State:}
Since SP is equivalent to averaging the mean $ \bar{\boldsymbol{h}}_* $ and the square root of pseudo-variance $ \check{\boldsymbol{h}}_* $, then $ \bar{\boldsymbol{h}}_*=\mathbb{E} _{n\in N}\bar{\boldsymbol{H}}_{*,n} $ and $ \check{\boldsymbol{h}}_*=\mathbb{E} _{n\in N}\check{\boldsymbol{H}}_{*,n} $.
Based on the assumption for spectral, the projection of each chunk also applies to the overall information.
Thus, the resued utterance-level loss $ \mathcal{L}_{\mathrm{rul}} $ is defined in Eq. \ref{eq:rul_loss}.

\noindent
\begin{equation}
  \small
  \label{eq:rul_loss}
  \mathcal{L} _{\mathrm{rul}}=\mathrm{MSE}\left( \boldsymbol{W}_m\bar{\boldsymbol{h}}_{s},\bar{\boldsymbol{h}}_{t} \right) + \mathrm{MSE}\left( \boldsymbol{W}_v\check{\boldsymbol{h}}_{s},\check{\boldsymbol{h}}_{t} \right)
\end{equation}
\noindent

\noindent
\textbf{Overall Loss:}
Combined with vanilla KD loss $ \mathcal{L} _{\mathrm{KD}} $ in Eq. \ref{eq:kd_loss} with Kullback-Leibler divergence (KLD), the overall loss $ \mathcal{L} _{\mathrm{rKD}} $ is defined in Eq. \ref{eq:rkd_loss}, where $ \eta $, $ \xi $ are hyperparameters.

\noindent
\begin{equation}
  \small
  \label{eq:kd_loss}
  \mathcal{L} _{\mathrm{KD}}=\mathrm{KLD}\left( f_t\left( x \right) /\tau \parallel f_s\left( x \right) /\tau \right)
\end{equation}
\noindent

\noindent
\begin{equation}
  \small
  \label{eq:rkd_loss}
  \mathcal{L} _{\mathrm{rKD}}=\mathcal{L} _{\mathrm{KD}}+\eta \cdot \mathcal{L} _{\mathrm{rfl}}+\xi \cdot \mathcal{L} _{\mathrm{rul}}
\end{equation}
\noindent

\section{Experiment}
\label{sec:experiment}

\subsection{Experimental Setup}

\textbf{Dataset:}
Both Urbansound8k (US8k) and ESC-50 (ESC50) \cite{piczak2015esc} use the last fold for evaluation and the others for training.
audioMNIST (AMnst) \cite{becker2018interpreting} uses the first 54 speakers for training and the last 6 speakers for evaluation.

\noindent
\textbf{Models:}
For the backbone model, there are two sets of settings for the teacher model (T.) and student model (S.):
\Circle{1}
Resnet-34 (res34) and Resnet-18 (res18).
\Circle{2}
WRN-40-2 (wrn40) and WRN-16-1 (wrn16).
The pooling layer of the backend model is changed to SP.

\noindent
\textbf{Data Driven:}
The teacher and student models are trained on the training data.
Then, train the student model to match the soft logits of the teacher model using vanilla KD \cite{hinton2015distilling}.

\noindent
\textbf{Data Free:}
We first conduct experiments on DFKD using existing data-free learning frameworks ADI \cite{yin2020dreaming}.
But changed the data augmentation method for audio, including rolling (only for TID) and random cuts in the time dimension.
Finally experiment with our FRAMI framework.

\noindent
\textbf{Hyper-parameters:}
The 16,000 Hz audio features use a 40-dimensional Mel-spectrogram, the hop-size is 10ms, and the window-size is 25ms.
On TID data for US8k and ESC50, 2-second samples need to be generated for KD.
TD data for AMnst only takes 1 second, but has no $ \mathcal{T} _{\mathrm{finv}} $.

\begin{table}[t]
  \centering
  \footnotesize
  \setlength{\tabcolsep}{5pt}
  \caption{Comparison of different methods on 3 datasets.}
  \label{tab:datafree}
  \renewcommand{\arraystretch}{1.0}
  \begin{tabular}{lcc|ccc|cc}
    \hline
                     & \multicolumn{2}{c|}{\textbf{Method}} & \multicolumn{3}{c|}{\textbf{Data Driven}} & \multicolumn{2}{c}{\textbf{Data Free}}                                                                                      \\ \cline{2-8}
    \textbf{Acc(\%)} & T.                                   & S.                                        & \multicolumn{1}{c}{T.}                 & S.    & KD\cite{hinton2015distilling} & ADI\cite{yin2020dreaming} & \textbf{FRAMI} \\ \hline
    \textbf{US8k}    & res34                                & res18                                     & 80.05                                  & 76.70 & 75.63                         & 66.67                     & \textbf{79.93} \\
                     & wrn40                                & wrn16                                     & 77.65                                  & 72.63 & 75.27                         & 63.44                     & \textbf{78.14} \\ \hline
    \textbf{ESC50}   & res34                                & res18                                     & 68.25                                  & 62.50 & 67.00                         & 59.25                     & \textbf{67.75} \\
                     & wrn40                                & wrn16                                     & 64.50                                  & 60.75 & 63.50                         & 58.75                     & \textbf{63.50} \\ \hline
    \textbf{AMnst}   & res34                                & res18                                     & 99.90                                  & 99.53 & 99.87                         & 99.10                     & \textbf{99.80} \\
                     & wrn40                                & wrn16                                     & 99.83                                  & 99.47 & 99.80                         & 99.03                     & \textbf{99.73} \\ \hline
  \end{tabular}
\end{table}

\begin{table}[t]
  \centering
  \footnotesize
  \setlength{\tabcolsep}{18pt}
  \caption{Ablation study of FRAMI with or without $ \mathcal{T} _{\mathrm{finv}} $ and Reused for WRN-based models.}
  \label{tab:ablation}
  \renewcommand{\arraystretch}{1.0}
  \begin{tabular}{@{}lccc@{}}
    \toprule
    \textbf{Acc(\%)}                                              & \textbf{US8k}  & \textbf{ESC50} & \textbf{AMnst} \\ \midrule
    \textbf{FRAMI (full)}                                         & \textbf{78.14} & \textbf{63.50} & -              \\
    w/o $ \mathcal{T} _{\mathrm{finv}} $ (Eq. \ref{eq:finv_term}) & 76.82          & 63.25          & \textbf{99.73} \\
    w/o Reused ($ \S $ \ref{subsec:reusedKD})                     & 77.54          & 62.75          & -              \\
    w/o $ \mathcal{T} _{\mathrm{finv}} $ w/o Reused               & 76.58          & 63.00          & 99.67          \\ \bottomrule
  \end{tabular}
  \vspace{-1em}
\end{table}

\subsection{Evaluation Results}

\textbf{Overall Evaluation:}
Table \ref{tab:datafree} compares the accuracy (Acc) of different KD methods with or without access to the original data.
Taking the ADI model as the baseline model, our proposed method achieves substantial accuracy improvements of 21.5\%, 11.2\% and 0.7\% relatively on US8k, ESC50 and AMnst, respectively.
The ADI model is slightly less accurate than the vanilla KD model.
But surprisingly, on US8k and ESC50, our approach achieves even better results than the data-driven KD method and is far superior to the data-driven student model.
Moreover, our method surpasses the teacher in the WRN-based model on the US8K.
In our analysis, the model inversion not only restores the real-like sample, but also mingles various characteristics to create more samples with richer features.
This is similar to the idea of \cite{gazneli2022end}, which merges features from multiple samples under the same label to achieve additional accuracy gains.

\noindent
\textbf{Ablation Studies:}
Table \ref{tab:ablation} shows the effect of the individual feature invariance term ($ \mathcal{T} _{\mathrm{finv}} $) and the reused teacher backend for KD (Reused), or their combination on the student accuracy under each dataset.
$ \mathcal{T} _{\mathrm{finv}} $ has a greater impact on US8k, but not on ESC50.
From our observations, the sample characteristics in the US8k dataset are stable and persistent, which is more consistent with the properties of TID.
However, most of the sample characteristics of the ESC50 data are more transient and more consistent with TD.
Hence, $ \mathcal{T} _{\mathrm{finv}} $ is more appropriate for the TID dataset.
The accuracy improves with the addition of Reuse methods, but is even better when combined with $ \mathcal{T} _{\mathrm{finv}} $.
Since $ \mathcal{T} _{\mathrm{finv}} $ creates more feature-rich samples, more valuable information can be captured at the frame-level, thereby improving accuracy.
In addition, this method also achieves a certain improvement in accuracy on AMnst.

\noindent
\textbf{Generated Samples:}
Fig. \ref{fig:adi_sample} to \ref{fig:cmi_sample} shows ``dog\_bark'' samples trained on US8k.
It can be seen that the spectrogram sample generated by the conventional method contains only a little feature matching the barking dog.
Whereas, by adding $ \mathcal{T} _{\mathrm{finv}} $, the generated samples have a more balanced feature distribution.
Fig. \ref{fig:0} to \ref{fig:9} show the generated AMnst samples from 0 to 9.
Furthermore, the synthesized samples are clearly audible to the human ear.

\begin{figure}[t]
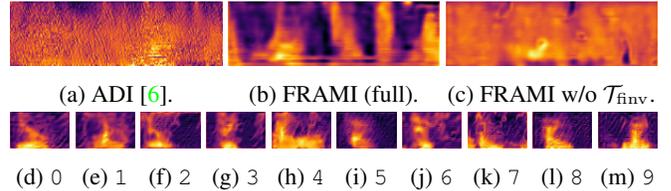

  \centering
  \caption{Spectrogram and audio of the generated samples by WRN-based models \textcolor[rgb]{0.6,0.6,0.6}{(\textit{click on the figure to hear the sound})}.}
  \foreach \i\j in {
  adi/ADI \cite{yin2020dreaming}.,
  frami/FRAMI (full).,
  cmi/{FRAMI w/o $ \mathcal{T} _{\mathrm{finv}} $.}
  }
  {\FigDog{\i}{\j}}

  \foreach \i in {0,...,9}
    {\FigDigit{\i}}



  \vspace{-1.5em}
\end{figure}

\section{Conclusions}
\label{sec:conclusion}

In this paper, we propose FRAMI, a framework for data-free knowledge distillation for general sound classification tasks.
We design a feature invariance contrastive inversion to ensure the feature richness of the generated samples, avoiding the problem of sparse audio features produced by traditional methods.
In knowledge distillation, the student model uses these feature-rich samples to mimic the teacher model at a deeper level by simultaneously learning the hidden states before and after the statistics pooling layer.
Experimental results on Urbansound8k, ESC-50, and audioMNIST demonstrate that both methods, alone or in combination, improve the accuracy of the student model.
Although this is a simple, preliminary exploration, we validate the feasibility of data-free knowledge distillation in general sound classification and are convinced that it will be extended to more audio models and more audio scenarios.

\section{Acknowledgement}
\label{sec:ack}

This paper is supported by the Key Research and Development Program of Guangdong Province under grant No.2021B\protect\\0101400003.
Corresponding author is Jing Xiao from Ping An Technology (Shenzhen) Co., Ltd (xiaojing661@ping-an.com.cn).


\bibliographystyle{IEEEbib}
\bibliography{refs}

\end{document}